\documentclass{ptptex}
\usepackage{wrapft}
\usepackage{graphicx}

\newcommand{\ket}[1]{| {#1} \rangle}     
\newcommand{\maru}[1]{\breve{#1}} 
\newcommand{\wtilde}[1]{\widetilde{#1}} 

\newcommand{\poon}[1]{\stackrel{\rightharpoonup} {#1}} 
\def\bsub{\begin{subequations}}
\def\esub{\end{subequations}}
\def\beq{\begin{eqnarray}}
\def\eeq{\end{eqnarray}}
\def\bsub{\begin{subequations}}
\def\esub{\end{subequations}}
\def\b{\begin{equation}}
\def\bs{\begin{split}}
\def\es{\end{split}}
\def\e{\end{equation}}

\begin{document}

\title{A note on the Lipkin model in arbitrary fermion number}

\author{Yasuhiko {\sc Tsue}$^{1,2}$, {Constan\c{c}a {\sc Provid\^encia}}$^{1}$, {Jo\~ao da {\sc Provid\^encia}}$^{1}$ and {Masatoshi {\sc Yamamura}}$^{1,3}$
}

\inst{$^{1}${CFisUC, Departamento de F\'{i}sica, Universidade de Coimbra, 3004-516 Coimbra, 
Portugal}\\
$^{2}${Physics Division, Faculty of Science, Kochi University, Kochi 780-8520, Japan}\\
$^{3}${Department of Pure and Applied Physics, 
Faculty of Engineering Science,\\
Kansai University, Suita 564-8680, Japan}\\
}

\abst{
A possible form of the Lipkin model obeying the $su(6)$-algebra is presented. 
It is a natural generalization from the idea for the $su(4)$-algebra recently 
proposed by the present authors. 
All the relation appearing in the present form can be expressed in terms of the spherical tensors 
in the $su(2)$-algebras. 
For specifying the linearly independent basis completely, twenty parameters are introduced. 
It is concluded that, in these parameters, 
the ten denote the quantum numbers coming from the eigenvalues of some hermitian operators. 
The five in these ten determine the minimum weight state. 
}


\maketitle

\section{Introduction}

This paper is a continuation of two papers, recently, published by the present authors \cite{1,2}. 
Hereafter, these two will be referred to as (I) and (II), respectively. 
In these papers, we treated the Lipkin model \cite{3,4} with arbitrary single-particle levels and 
fermion number. 
In (I), mainly an idea of how to construct the minimum weight state, which is the starting point of the algebraic approach, was proposed. 
In (II), we discussed how to express the linearly independent basis built on a chosen minimum weight state. 
Present paper aims mainly at supplementing the results of (II) with the 
discussion on the $su(6)$-algebra, which we promised in (II).

First, let us consider the Lipkin model obeying the $su(n)$-algebra for the case with $n=2m$ in rather general framework. 
Here, $m$ denotes integer. 
Since the total fermion number $N$ is a constant of motion, we omit the discussion on $N$. 
Our present argument is restricted to the case with even integer $n$ with $n=4,\ 6,\cdots$, i.e., $n=2m$ with $m=2,\ 3,\cdots$. 
If the generators in the $su(n)$-algebra are expressed appropriately, we can show that this model includes 
$m$ $su(2)$-subalgebras. 
This point has been shown in (II) and the generators are given in the relation (II.2.3). 
Therefore, as the total sum, we can define the $su(2)$-algebra $({\wtilde S}_{\pm,0})$ which will play a central role 
in our approach. 
Of course, the generators in each $su(2)$-algebra form a vector. 
Further, in (II), we showed that the $su(n)$-Lipkin model includes one $su(m)$-subalgebra, all generators in which are scalar for $({\wtilde S}_{\pm,0})$ 
and the remains form $m(m-1)/2$ vectors. 
They are given in the relations (II.2.9) and (II.2.10) $\sim$ (II.2.12), respectively. 
Generally, the minimum weight states in the $su(n)$-algebra are specified by $(n-1)$ quantum numbers. 
In the present case, we can decompose the number $(n-1)$ into two parts: 
\beq\label{1}
n-1=m+(m-1)\ . \qquad (n=2m)
\eeq 
As was already mentioned, the case with $n=2m$ includes $m$ $su(2)$-algebras and one $su(m)$-algebra. 
The first and the second term, $m$ and $(m-1)$, represent, respectively, the numbers of the 
quantum numbers related with $m$ $su(2)$-algebras and of one $su(m)$-algebra for the minimum weight state.

Next, let us consider the orthogonal set constructed by operating ``certain operators" on any minimum weight state. 
We will call them as the excited state generating operators. 
They should be expressed in terms of $n(n-1)/2$ quantum numbers coming from the relation 
\beq\label{2}
\frac{1}{2}\left(\left(n^2-1\right)-\left(n-1\right)\right)
=\frac{1}{2}n(n-1)
=m(2m-1)\ . \qquad (n=2m)
\eeq
The number $n(n-1)/2$ is equal to that of the $su(n)$-generators with the types ${\wtilde S}^p(n)$ $(p=1,\ 2,\cdots ,\ n-1)$ and 
${\wtilde S}_q^p(n)$ $(p>q=1,\ 2,\cdots ,\ n-2)$: 
$(n-1)+(n-1)(n-2)/2=n(n-1)/2$. 
We will call them as the raising operators and their hermitian conjugates as the lowering operators. 
The definition of ${\wtilde S}^p(n)$ and ${\wtilde S}_q^p(n)$ has been given in the relation (I.2.2). 
We know that there exist $m$ $su(2)$-subalgebras and one $su(m)$-algebra and, then, in the excited state generating operators, $m$ and $m(m-1)/2$ operators 
are related to these subalgebras, respectively.  
Therefore, we must investigate the remain, the number of which is given by
\beq\label{3}
m(2m-1)-\left(m+\frac{1}{2}m(m-1)\right)=3\cdot \frac{1}{2}m(m-1)\ . 
\eeq
It should be noted that the number $m(m-1)/2$ is just equal to that of the vector operators 
for $({\wtilde S}_{\pm,0})$, which are presented in (II). 
With the use of these three types of the operators, we can expect to obtain a possible idea for constructing the excited state generating operators. 
Further, we notice that the relation (\ref{3}) can be decomposed into 
\beq\label{4}
3\cdot \frac{1}{2}m(m-1)=\frac{1}{2}m(m-1)+2\cdot \frac{1}{2}m(m-1)\ . 
\eeq
First term corresponds to the number of the raising or the lowering operators in the $su(m)$-subalgebra. 
It is well known that we can construct a tensor operator with two parameters in terms of a vector, for example, the solid harmonics 
${\mathcal Y}_{l,l_0}=r^l Y_{ll_0}(\theta \phi)$ is constructed in terms of the position vector ($x=r\sin \theta \cos \phi,\ 
y=r\sin \theta \sin \phi,\ z=r\cos \phi)$. 
In our case, there exist $m(m-1)/2$ vectors and each gives us a tensor operator in the $su(2)$-subalgebras, which is 
specified by two parameters. 
This argument leads us to the following: 
Second term represents the total number of the parameters contained in the operators 
which are built by $m(m-1)/2$ vectors. 
On the other hand, first term $m(m-1)/2$ corresponds to the number of 
the lowering operators in the $su(m)$-subalgebra. 
A merit of our idea may be as follows: 
All generators of the $su(n)$-Lipkin model are expressed in terms of the spherical tensors for $({\wtilde S}_{\pm,0})$ and, then, 
we can apply the technique of the angular momentum coupling rule.

As a simple example of the Lipkin model, we will give a brief summary for the case with $n=4$, i.e., $m=2$, which has been 
discussed in (II), but the form in this paper is a little bit different from that of (II). 
The generators in the present two $su(2)$-subalgebras are copied from the relation (II.5.9): 
\setcounter{equation}{3}
\bsub\label{4ab}
\beq
& &{\wtilde S}_+(1)={\wtilde S}_2^3\ , \qquad
{\wtilde S}_-(1)={\wtilde S}_3^2\ , \qquad
{\wtilde S}_0(1)=\frac{1}{2}\left({\wtilde S}_3^3-{\wtilde S}_2^2\right)\ , 
\label{4a}\\
& &{\wtilde S}_+(2)={\wtilde S}^1\ , \qquad
{\wtilde S}_-(2)={\wtilde S}_1\ , \qquad 
{\wtilde S}_0(2)=\frac{1}{2}{\wtilde S}_1^1\ . 
\label{4b}
\eeq
\esub
The sum is given by 
\beq\label{5}
{\wtilde S}_{\pm,0}={\wtilde S}_{\pm,0}(1)+{\wtilde S}_{\pm,0}(2)\ . 
\eeq
The generators in the $su(m=2)$-subalgebra are expressed as 
\beq\label{6}
{\wtilde R}_+={\wtilde S}_1^3+{\wtilde S}^2\ , \qquad
{\wtilde R}_-={\wtilde S}_3^1+{\wtilde S}_2\ , \qquad
{\wtilde R}_0=\frac{1}{2}\left({\wtilde S}_3^3+{\wtilde S}_2^2-{\wtilde S}_1^1\right)\ . 
\eeq
Since ${\wtilde R}_{\pm,0}$ are scalars for $({\wtilde S}_{\pm,0})$, we have the relation 
\beq\label{7}
\left[\ {\rm any\ of}\ {\wtilde S}_{\pm,0}\ , \ {\rm any\ of}\ {\wtilde R}_{\pm,0}\ \right]=0\ . 
\eeq
The vector operators for $({\wtilde S}_{\pm,0})$ are given as 
\bsub\label{8}
\beq
& &{\wtilde R}^{1,+1}=-{\wtilde S}^3\ , \qquad
{\wtilde R}^{1,0}=\frac{1}{\sqrt{2}}\left({\wtilde S}_1^3-{\wtilde S}^2\right)\ , \qquad
{\wtilde R}^{1,-1}={\wtilde S}_1^2\ , 
\label{8a}\\
& &{\wtilde R}_{1,+1}=-{\wtilde S}_3\ , \qquad
{\wtilde R}_{1,0}=\frac{1}{\sqrt{2}}\left({\wtilde S}_3^1-{\wtilde S}_2\right)\ , \qquad
{\wtilde R}_{1,-1}={\wtilde S}_2^1\ , 
\label{8b}
\eeq
\esub
The relations (\ref{6})$\sim$ (\ref{8}) are copied from the relations (II.5.11)$\sim$ (II.5.14). 
The vectors ${\wtilde R}_{\pm,0}$ satisfy 
\beq\label{9}
\left[\ {\wtilde R}_+\ , \ {\wtilde R}^{1,\nu}\ \right] =0\ , \qquad
\left[\ {\wtilde R}_0\ , \ {\wtilde R}^{1,\nu}\ \right]={\wtilde R}^{1,\nu}\ . \qquad (\nu=\pm,\ 0)
\eeq

The minimum weight state is expressed as $\ket{\rho,\sigma^1,\sigma^2}$, where the eigenvalues of ${\wtilde R}_0$, 
${\wtilde S}_0(1)$ and ${\wtilde S}_0(2)$ are denoted as $-\rho$, $-\sigma^1$ and $-\sigma^2$, respectively. 
The explicit form of $\ket{\rho,\sigma^1,\sigma^2}$ is given in (I). 
Then, the eigenstate of $({\wtilde {\mib R}}^2,{\wtilde R}_0)$, $({\wtilde {\mib S}}(1)^2, {\wtilde S}_0(1))$ 
and $({\wtilde {\mib S}}(2)^2, {\wtilde S}_0(2))$ with the eigenvalues 
$(\rho(\rho+1), \rho_0)$, $(\sigma^1(\sigma^1+1), \sigma_0^1)$ and $(\sigma^2(\sigma^2+1), \sigma_0^2)$ can be 
expressed as 
\bsub\label{10}
\beq
& &
\ket{\rho_0,\sigma_0^1,\sigma_0^2;\rho,\sigma^1,\sigma^2}
=f_{\rho\rho_0}f_{\sigma^1\sigma_0^1}f_{\sigma^2\sigma_0^2}{\wtilde P}^{\rho\rho_0,\sigma^1\sigma_0^1,\sigma^2\sigma_0^2}\ket{\rho,\sigma^1,\sigma^2}\ , 
\label{10a}\\
& &
{\wtilde P}^{\rho\rho_0,\sigma^1\sigma_0^1,\sigma^2\sigma_0^2}=\left({\wtilde R}_+\right)^{\rho+\rho_0}
\left({\wtilde S}_+(1)\right)^{\sigma^1+\sigma_0^1}\left({\wtilde S}_+(2)\right)^{\sigma^2+\sigma_0^2}\ , 
\label{10b}
\eeq
\esub
\beq\label{11}
& &f_{\tau\tau_0}=\sqrt{\frac{(\tau-\tau_0)!}{(2\tau)!(\tau+\tau_0)!}}\ . \qquad (\tau=\rho,\ \sigma^1,\ \sigma^2)
\qquad\qquad\qquad
\eeq
The eigenstate of $({\wtilde {\mib S}}^2, {\wtilde S}_0)$ with the eigenvalue $(\sigma(\sigma+1),\sigma_0)$ is 
given by 
\beq\label{12}
\ket{\rho\rho_0,\sigma^1,\sigma^2,\sigma\sigma_0}
=\sum_{\sigma_0^1 \sigma_0^2}\langle \sigma^1\sigma_0^1\sigma^2\sigma_0^2\ket{\sigma\sigma_0}
\ket{\rho_0,\sigma_0^1,\sigma_0^2;\rho,\sigma^1,\sigma^2}\ . 
\eeq
We are investigating the case with $n=4$, i.e., $m=2$. 
Therefore, as can be seen in the relation (\ref{3}), further, we must search the operator 
characterized by three quantum numbers for obtaining the excited state generating operators. 
For this task, in (II), the following form is adopted: 
\beq\label{13}
{\wtilde R}^{l,l_0;\l,\lambda_0}=\left({\poon R}_-\right)^{l-\lambda_0}\left({\poon S}_-\right)^{l-l_0}
\left(-{\wtilde S}^3\right)^l\ . 
\eeq
Here, we adopted the notation for ${\wtilde O}$ and ${\wtilde A}$ in the form 
\beq\label{14}
\left({\poon O}\right)^n{\wtilde A}
=\underbrace{
\left[{\wtilde O}, \cdots ,\left[{\wtilde O},\left[{\wtilde O}\right.\right.\right.}_{n},{\wtilde A}\Bigl]\Bigl]\cdots \Bigl]\ .
\eeq
Since ${\poon S}_+\left(-{\wtilde S}^3\right)^l={\poon R}_+\left(-{\wtilde S}^3\right)^l=0$ and 
${\poon S}_0\left(-{\wtilde S}^3\right)^l={\poon R}_0\left(-{\wtilde S}^3\right)^l=l\left(-{\wtilde S}^3\right)^l$, 
${\wtilde R}^{l,l_0;l,\lambda_0}$ can be regarded as spherical tensor operator for the $su(2)$-algebras $({\wtilde S}_{\pm,0})$ and $({\wtilde R}_{\pm,0})$ with rank $l$. 
Operating ${\wtilde R}^{l,l_0;l,\lambda_0}$ on the state (\ref{12}) and applying the angular momentum coupling rule, we have 
\beq\label{15}
& &\ket{l,rr_0,ss_0,\sigma;\rho\sigma^1\sigma^2}
=
\sum_{\lambda_0 l_0}\langle l\lambda_0\rho\rho_0 \ket{rr_0}\langle ll_0\sigma\sigma_0 \ket{ss_0}
{\wtilde R}^{l,l_0;l,\lambda_0}
\ket{\rho\rho_0,\sigma^1,\sigma^2,\sigma\sigma_0}\ . 
\eeq
However, the set composed by the states (\ref{15}) cannot be regarded as orthogonal. 
The reasons are as follows: 
(1) the symbol $l$ is not a quantum number, but a parameter, the value of which is given from the outside 
and (2) the operator ${\wtilde R}^{l,l_0;l,\lambda_0}$ contains the degrees of freedom which do not contain in the states 
$\ket{\rho\rho_0,\sigma^1,\sigma^2,\sigma\sigma_0}$, 
for example, such as $-{\wtilde S}^3$. 
Therefore, they form the linearly independent basis and by appropriate method, for example, by the Schmidt method, we must 
construct the orthogonal set. 
In the sense mentioned above, 
the excited state generating operator is given by 
\beq\label{16}
{\wtilde R}^{l,l_0;l,\lambda_0}\cdot {\wtilde P}^{\rho\rho_0,\sigma^1\sigma_0^1,\sigma^2\sigma_0^2}
&=&
\left({\poon R}_-\right)^{l-\lambda_0}\left({\poon S}_-\right)^{l-l_0}\left(-{\wtilde S}^3\right)^l
\nonumber\\
& &
\times \left({\wtilde R}_+\right)^{\rho+\rho_0}\left({\wtilde S}_+(1)\right)^{\sigma^1+\sigma_0^1}
\left({\wtilde S}_+(2)\right)^{\sigma^2+\sigma_0^2}\ . 
\eeq

The above is the summary of (II) for the case with $n=4$, i.e., $m=2$. 
In the preparation for the discussion on the case with $n=6$, i.e., $m=3$, 
we rewrite the relation (\ref{13}) in the form slightly different from the 
original. 
The form $({\poon S}_-)^{l-l_0}(-{\wtilde S}^3)^l$ can be rewritten as 
\beq\label{17}
\left({\poon S}_-\right)^{l-l_0}\left(-{\wtilde S}^3\right)^l
=\left({\poon S}_-\right)^{l-l_0}\left({\wtilde R}^{1,+1}\right)^l
={\wtilde Z}^{l,l_0}\ . 
\eeq
Here, ${\wtilde Z}^{l,l_0}$ is of the form 
\beq\label{18}
& &
{\wtilde Z}^{l,l_0}=\sqrt{\frac{l!}{(2l-1)!!}}\sum_{\lambda}
\left(\frac{1}{\sqrt{2}}\right)^{l-\lambda}
\frac{\sqrt{(l+l_0)!(l-l_0)!}}{\left(\frac{l+l_0-\lambda}{2}\right)!\lambda!\left(\frac{l-l_0-\lambda}{2}\right)!}
\nonumber\\
& &\qquad\qquad\qquad\qquad\qquad\qquad
\times
\left({\wtilde R}^{1,+1}\right)^{\frac{l+l_0-\lambda}{2}}
\left({\wtilde R}^{1,0}\right)^\lambda
\left({\wtilde R}^{1,-1}\right)^{\frac{l-l_0-\lambda}{2}} .
\eeq
The sum for $\lambda$ obeys the following condition : 
if $|l-l_0|=$even or odd, 
$\lambda$ cannot be odd or even, respectively. 
The operator ${\wtilde Z}^{l,l_0}$ is a tensor with rank $l$ for $({\wtilde S}_{\pm,0})$: 
\beq\label{19}
& &{\poon S}_{\pm}{\wtilde Z}^{l,l_0}=\sqrt{(l\mp l_0)(l\pm l_0+1)}{\wtilde Z}^{l,l_0\pm 1}\ , 
\nonumber\\
& &{\poon S}_0{\wtilde Z}^{l,l_0}=l_0{\wtilde Z}^{l,l_0}\ . 
\eeq
In the case that $({\wtilde R}^{1,\pm 1}, {\wtilde R}^{1,0})$ is a position vector 
$(r_{\pm 1}=\mp(x\pm iy)/\sqrt{2}, r_0=z)$, ${\wtilde Z}^{l,l_0}$ is reduced to the solid harmonics: 
\beq\label{20}
{\wtilde Z}^{l,l_0} \longrightarrow 
\sqrt{\frac{4\pi l!}{(2l+1)!!}}{\mathcal Y}_{l,l_0}\ , \qquad
\left( {\mathcal Y}_{l,l_0}=r^l Y_{ll_0}(\theta \phi)\right)\ . 
\eeq
An important property of ${\wtilde Z}^{l,l_0}$ is as follows: 
\beq\label{21}
{\poon R}_+{\wtilde Z}^{l,l_0}=0\ , \qquad
{\poon R}_0{\wtilde Z}^{l,l_0}=l{\wtilde Z}^{l,l_0}\ . 
\eeq
With the use of the relation (\ref{9}), we are able to obtain the property (\ref{21}).

Under the above preparation, we will investigate the case with $n=6$, i.e., $m=3$. 
This case includes three $su(2)$-subalgebras in the form 
\bsub\label{22}
\beq
& &{\wtilde S}_+(1)={\wtilde S}_4^5\ , \qquad
{\wtilde S}_-(1)={\wtilde S}_5^4\ , \qquad
{\wtilde S}_0(1)=\frac{1}{2}\left({\wtilde S}_5^5 -{\wtilde S}_4^4\right)\ , 
\label{22a}\\
& &{\wtilde S}_+(2)={\wtilde S}_2^3\ , \qquad
{\wtilde S}_-(2)={\wtilde S}_3^2\ , \qquad
{\wtilde S}_0(2)=\frac{1}{2}\left({\wtilde S}_3^3 -{\wtilde S}_2^2\right)\ , 
\label{22b}\\
& &{\wtilde S}_+(3)={\wtilde S}^1\ , \qquad
{\wtilde S}_-(3)={\wtilde S}_1\ , \qquad
{\wtilde S}_0(3)=\frac{1}{2}{\wtilde S}_1^1 \ . 
\label{22c}
\eeq
\esub
The total sum of the above is denoted as 
\beq\label{23}
{\wtilde S}_{\pm,0}={\wtilde S}_{\pm,0}(1)+{\wtilde S}_{\pm,0}(2)+{\wtilde S}_{\pm, 0}(3)\ . 
\eeq
The eight generators in the $su(m=3)$-subalgebra, which are scalars for $({\wtilde S}_{\pm,0})$, are written down as follows: 
\bsub\label{24}
\beq
& &{\wtilde R}_+={\wtilde S}_3^5+{\wtilde S}_2^4\ , \qquad
{\wtilde R}_-={\wtilde S}_5^3+{\wtilde S}_4^2\ , \qquad
{\wtilde R}_0=\frac{1}{2}\left({\wtilde S}_5^5 +{\wtilde S}_4^4-{\wtilde S}_3^3-{\wtilde S}_2^2\right)\ , 
\label{24a}\\
& &{\wtilde R}^{\frac{1}{2},\frac{1}{2}}={\wtilde S}_1^5+{\wtilde S}^4\ , \qquad
{\wtilde R}_{\frac{1}{2},\frac{1}{2}}={\wtilde S}_5^1+{\wtilde S}_4\ ,
\nonumber\\
& &{\wtilde R}^{\frac{1}{2},-\frac{1}{2}}={\wtilde S}_1^3+{\wtilde S}^2\ , \qquad
{\wtilde R}_{\frac{1}{2},-\frac{1}{2}}={\wtilde S}_3^1+{\wtilde S}_2\ ,
\label{24b}\\
& &{\wtilde R}=\frac{1}{2}\left({\wtilde S}_5^5 +{\wtilde S}_4^4+{\wtilde S}_3^3+{\wtilde S}_2^2\right)-{\wtilde S}_1^1\ . 
\label{24c}
\eeq
\esub
The set (\ref{24a}) forms the $su(2)$-algebra and $({\wtilde R}^{\frac{1}{2},\frac{1}{2}}, {\wtilde R}^{\frac{1}{2},-\frac{1}{2}})$ and 
$(-{\wtilde R}_{\frac{1}{2},-\frac{1}{2}}, {\wtilde R}_{\frac{1}{2},\frac{1}{2}})$ are spinors for $({\wtilde R}_{\pm,0})$. 
The present case includes three vectors for $({\wtilde S}_{\pm,0})$: 
\bsub\label{25}
\beq
& &{\wtilde R}^{1,+1}(1)=-{\wtilde S}^5\ , \qquad
{\wtilde R}^{1,0}(1)=\frac{1}{\sqrt{2}}\left({\wtilde S}_1^5-{\wtilde S}^4\right)\ , \qquad
{\wtilde R}^{1,-1}(1)={\wtilde S}_1^4\ , 
\label{25a}\\
& &{\wtilde R}^{1,+1}(2)=-{\wtilde S}_2^5\ , \qquad
{\wtilde R}^{1,0}(2)=\frac{1}{\sqrt{2}}\left({\wtilde S}_3^5-{\wtilde S}_2^4\right)\ , \qquad
{\wtilde R}^{1,-1}(2)={\wtilde S}_3^4\ ,
\label{25b}\\
& &{\wtilde R}^{1,+1}(3)=-{\wtilde S}^3\ , \qquad
{\wtilde R}^{1,0}(3)=\frac{1}{\sqrt{2}}\left({\wtilde S}_1^3-{\wtilde S}^2\right)\ , \qquad
{\wtilde R}^{1,-1}(3)={\wtilde S}_1^2\ . 
\label{25c}
\eeq
\esub
The hermitian conjugates of the vectors are omitted to give. 
The expressions (\ref{22}), (\ref{24}) and (\ref{25}) are obtained by putting $m=3$ in the 
relations (II.2.3) and (II.2.9)$\sim$(II.2.12). 
For $\nu=\pm 1,\ 0$, the vector 
$({\wtilde R}^{1,\nu}(1))$ satisfies the relation 
\bsub\label{26}
\beq
& &{\poon R}_+{\wtilde R}^{1,\nu}(1)={\poon R}{}^{\frac{1}{2},\frac{1}{2}}{\wtilde R}^{1,\nu}(1)={\poon R}{}^{\frac{1}{2},-\frac{1}{2}}{\wtilde R}^{1,\nu}(1)=0\ , 
\label{26a}\\
& &{\poon R}_0{\wtilde R}^{1,\nu}(1)=\frac{1}{2}{\wtilde R}^{1,\nu}(1)\ , \qquad
{\poon R}{\wtilde R}^{1,\nu}(1)=\frac{3}{2}{\wtilde R}^{1,\nu}(1)\ . 
\label{26b}
\eeq
\esub
For $\nu=\pm 1,\ 0$ and $k=2,\ 3$, two vectors $({\wtilde R}^{1,\nu}(k))$ obey the relation 
\bsub\label{27}
\beq
& &{\poon R}_+{\wtilde R}^{1,\nu}(k)=\delta_{k,3}{\wtilde R}^{1,\nu}(k)\ , \quad
{\poon R}{}^{\frac{1}{2},\frac{1}{2}}{\wtilde R}^{1,\nu}(k)=0\ , \quad
{\poon R}{}^{\frac{1}{2},-\frac{1}{2}}{\wtilde R}^{1,\nu}(k)=-\delta_{k,2}{\wtilde R}^{1,\nu}(k)\ , \qquad
\label{27a}\\
& &{\poon R}_0{\wtilde R}^{1,\nu}(k)=\left(\delta_{k,2}-\delta_{k,3}\cdot\frac{1}{2}\right){\wtilde R}^{1,\nu}(k)\ , \quad
{\poon R}{\wtilde R}^{1,\nu}(k)=\delta_{k,3}\cdot\frac{3}{2}{\wtilde R}^{1,\nu}(k)\ . 
\label{27b}
\eeq
\esub
The above are the relations for our present case $(n=6)$. 
However, ${\wtilde R}^{1,\nu}(2)$ and ${\wtilde R}^{1,\nu'}(3)$ do not commute mutually, but, 
they commute with ${\wtilde R}^{1,\nu''}(1)$. 
We will treat the case with $n=6$ as a natural generalization from the case with $n=4$.

Now, in parallel with the $su(4)$-Lipkin model, we are possible to give our scheme for obtaining the linearly independent basis for the $su(6)$-Lipkin model. 
In order to avoid unnecessary complication, we will not apply the angular momentum coupling rule, together with the associating numerical factors, 
for example, such as the form (\ref{11}). 
The minimum weight state is expressed in the form $\ket{\rho,\rho^1,\sigma^1,\sigma^2,\sigma^3}$, 
where $\rho$, $\rho^1$, $\sigma^1$, $\sigma^2$ and $\sigma^3$ denote the eigenvalues of 
$-{\wtilde R}$, $-{\wtilde R}_0$, $-{\wtilde S}_0(1)$, $-{\wtilde S}_0(2)$ and $-{\wtilde S}_0(3)$ defined in the relations (\ref{24}) and (\ref{22}), respectively. 
As a possible extension of ${\wtilde P}^{\rho\rho_0,\sigma^1\sigma_0^1,\sigma^2\sigma_0^2}$ shown in the relation (\ref{10b}), 
we introduce the following operator: 
\beq\label{28}
& &{\wtilde P}^{\mu\mu_0,\rho^1\rho_0^1,\sigma^1\sigma_0^1,\sigma^2\sigma_0^2,\sigma^3\sigma_0^3}
\nonumber\\
&=&
\!\!\left({\wtilde R}^{\frac{1}{2},\frac{1}{2}}\right)^{\mu+\mu_0}\left({\wtilde R}^{\frac{1}{2},-\frac{1}{2}}\right)^{\mu-\mu_0}
\left({\wtilde R}_+\right)^{\rho^1+\rho_0^1}\left({\wtilde S}_+(1)\right)^{\sigma^1+\sigma_0^1}\left({\wtilde S}_+(2)\right)^{\sigma^2+\sigma_0^2}
\left({\wtilde S}_+(3)\right)^{\sigma^3+\sigma_0^3} . \quad
\eeq
With the use of the operator (\ref{28}), we define the state 
\beq\label{29}
& &\ket{\mu\mu_0,\rho_0^1,\sigma_0^1,\sigma_0^2,\sigma_0^3;\rho,\rho^1,\sigma^1,\sigma^2,\sigma^3}
=
{\wtilde P}^{\mu\mu_0,\rho^1\rho_0^1,\sigma^1\sigma_0^1,\sigma^2\sigma_0^2,\sigma^3\sigma_0^3}\ket{\rho,\rho^1,\sigma^1,\sigma^2,\sigma^3}\ . 
\eeq
The state (\ref{29}) corresponds to the state (\ref{10a}) and is expressed in 
terms of eleven parameters, some of which play the role of quantum numbers.

The above is first step in our approach to the case with $n=6$. 
The linearly independent basis of our present case should be expressed totally in terms of twenty parameters. 
Therefore, we must, further, investigate how to consider nine parameters which may be related with three vectors. 
For this task, we extend the idea adopted in the case with $n=4$. 
This case starts in the relation (\ref{9}). 
The relation (\ref{9}) for the vector (\ref{17}) with (\ref{18}) for the definition of ${\wtilde Z}^{l,l_0}$ leads us to the relation (\ref{21}) 
for the raising operator ${\wtilde R}_+$ and the hermitian operator ${\wtilde R}_0$. 
These two correspond to $({\wtilde R}_+,{\wtilde R}^{\frac{1}{2},\frac{1}{2}}, {\wtilde R}^{\frac{1}{2},-\frac{1}{2}})$ and $({\wtilde R}_0, {\wtilde R})$, 
respectively for the present case. 
As can be seen in the relation (\ref{27a}), the operation of ${\poon R}_+$ and ${\poon R}^{\frac{1}{2},\pm\frac{1}{2}}$ on the vectors labeled $k=2,\ 3$ do not vanish. 
Then, if we try to use the form extended from the relation (\ref{9}), we must introduce new vectors. 
As the candidate of these new, the following are possible to use: 
\bsub\label{30}
\beq
& &{\maru R}{}^{1,\nu}(1)={\wtilde R}^{1,\nu}(1)\ , \quad (\nu=\pm 1,\ 0)
\label{30a}\\
& &{\maru R}{}^{1,1}(k)=\frac{1}{\sqrt{2}}\left({\wtilde R}^{1,1}(1){\wtilde R}^{1,0}(k)-{\wtilde R}^{1,1}(k){\wtilde R}^{1,0}(1)\right)\ , \nonumber\\
& &{\maru R}{}^{1,0}(k)=\frac{1}{\sqrt{2}}\left({\wtilde R}^{1,1}(1){\wtilde R}^{1,-1}(k)-{\wtilde R}^{1,1}(k){\wtilde R}^{1,-1}(1)\right)\ , \nonumber\\
& &{\maru R}{}^{1,-1}(k)=\frac{1}{\sqrt{2}}\left({\wtilde R}^{1,0}(1){\wtilde R}^{1,-1}(k)-{\wtilde R}^{1,0}(k){\wtilde R}^{1,-1}(1)\right)\ .
\label{30b}
\eeq
\esub
The vector (\ref{30b}) is derived by the formula $Z_\nu=\sum_{\lambda,\mu}\langle 1\lambda1\mu\ket{1\nu}X_{\lambda}Y_{\mu}$ 
for two vectors $X$ and $Y$. 
The vectors (\ref{30}) satisfy the following relation for $\nu=\pm 1,\ 0$ and $k=1,\ 2,\ 3$:
\bsub\label{31}
\beq
& &{\poon R}_+{\maru R}{}^{1,\nu}(k)={\poon R}^{\frac{1}{2},\pm\frac{1}{2}}{\maru R}{}^{1,\nu}(k)=0\ , 
\label{31a}\\
& &{\poon R}_0{\maru R}{}^{1,\nu}(k)=\frac{1}{2}\left(1+\delta_{k,2}-\delta_{k,3}\right){\maru R}{}^{1,\nu}(k)\ , 
\nonumber\\
& &{\poon R}{\maru R}{}^{1,\nu}(k)=\frac{3}{2}\left(1+\delta_{k,3}\right){\maru R}{}^{1,\nu}(k)\ . 
\label{31b}
\eeq
\esub
With the use of the vectors (\ref{31}), we can define the operator 
\beq\label{32}
{\maru Z}{}^{l^1l_0^1,l^2l_0^2,l^3l_0^3}={\wtilde Z}^{l^1l_0^1}(1){\wtilde Z}^{l^2l_0^2}(2){\wtilde Z}^{l^3l_0^3}(3)\ . 
\eeq
Here, ${\wtilde Z}^{l^kl_0^k}(k)$ $(k=1,\ 2,\ 3)$ is obtained by replacing $ll_0$ and 
${\wtilde R}^{1,\nu}$ in the relation (\ref{18}) with $l^kl_0^k$ and ${\maru R}{}^{1,\nu}(k)$. 
As was already mentioned, some pairs of $({\wtilde R}^{1,\nu}(k))$ do not commute and, therefore, the ordering of 
${\wtilde Z}^{l^kl_0^k}(k)$ in the definition (\ref{32}) should be fixed beforehand, for example, as is shown in the definition (\ref{32}). 
The operator (\ref{32}) satisfies 
\bsub\label{33}
\beq
& &{\poon R}_+{\maru Z}{}^{l^1l_0^1,l^2l_0^2,l^3l_0^3}={\poon R}^{\frac{1}{2},\pm\frac{1}{2}}{\maru Z}{}^{l^1l_0^1,l^2l_0^2,l^3l_0^3}=0\ , 
\label{33a}\\
& &{\poon R}_0{\maru Z}{}^{l^1l_0^1,l^2l_0^2,l^3l_0^3}=\frac{1}{2}\left(l^1+3l^2\right){\maru Z}{}^{l^1l_0^1,l^2l_0^2,l^3l_0^3}\ , 
\nonumber\\
& &{\poon R}{\maru Z}{}^{l^1l_0^1,l^2l_0^2,l^3l_0^3}=\frac{3}{2}\left(l^1+l^2+2l^3\right){\maru Z}{}^{l^1l_0^1,l^2l_0^2,l^3l_0^3}\ . 
\label{33b}
\eeq
\esub
We can see that the above is a natural extension from the relation (\ref{21}).

The operator (\ref{32}) is expressed in terms of six parameters $(l^1,l_0^1)$, $(l^2,l_0^2)$ and $(l^3,l_0^3)$. 
Then, in order to accomplish our task, we must search, further, three parameters. 
As can be seen in the relation (\ref{13}), the case with $n=4$ is completed by taking into account the lowering operator ${\wtilde R}_-$ in the form ${\poon R}_-$. 
The relation (\ref{33a}) tells us that the operation of the raising operators ${\wtilde R}_+$ and ${\wtilde R}^{\frac{1}{2},\pm\frac{1}{2}}$ makes the results vanish. 
It may be enough to consider three lowering operators ${\poon R}_-$ and $\mp {\poon R}_{\frac{1}{2},\pm\frac{1}{2}}$ on the operator (\ref{32}). 
First, we note that the operator (\ref{32}) is nothing but tensor specified by 
\beq\label{34}
l=l_0=\frac{1}{2}(l^1+3l^2)\ .
\eeq
Then, tensor operator specified by $l$ and $l_0$ $(l_0=-l,\ -l+1,\cdots ,\ l-1,\ l)$ can be given in the form 
\beq\label{35}
{\maru Z}_{ll_0}{}^{(l^1l_0^1,l^2l_0^2,l^3l_0^3)}=\left({\poon R}_-\right)^{l-l_0}
{\maru Z}{}^{l^1l_0^1,l^2l_0^2,l^3l_0^3}\ . 
\eeq
Next, we consider the operators $\mp {\wtilde R}_{\frac{1}{2},\mp \frac{1}{2}}$, which form the spinor for 
$({\wtilde R}_{\pm,0})$. 
As is well known, tensor operator specified by $\lambda$ and $\lambda_0$ $(\lambda_0=-\lambda,\ -\lambda+1,\cdots ,\ \lambda-1,\ \lambda)$ 
is constructed in the form 
\beq\label{36}
{\wtilde Y}_{\lambda\lambda_0}=\left(-{\tilde R}_{\frac{1}{2},-\frac{1}{2}}\right)^{\lambda+\lambda_0}\left({\wtilde R}_{\frac{1}{2},\frac{1}{2}}\right)^{\lambda-\lambda_0}\ . 
\eeq
In the case of applying the angular momentum couping rule, it may be convenient to attach the numerical factor $g_{\lambda\lambda_0}$: 
\beq\label{37}
g_{\lambda\lambda_0}=\sqrt{\frac{(2\lambda)!}{(\lambda+\lambda_0)!(\lambda-\lambda_0)!}}\ . 
\eeq
Then, we introduce the operator 
\beq\label{39}
{\poon Y}_{\lambda\lambda_0}=\left(-{\poon R}_{\frac{1}{2},-\frac{1}{2}}\right)^{\lambda+\lambda_0}\left({\poon R}_{\frac{1}{2},\frac{1}{2}}\right)^{\lambda-\lambda_0}\ .
\eeq
Product of the operators (\ref{35}) and (\ref{39}) gives us the operator with nine parameters:
\beq\label{40}
{\wtilde R}_{\lambda\lambda_0,ll_0}{}^{(l^1l_0^1,l^2l_0^2,l^3l_0^3)}
&=&
{\poon Y}_{\lambda\lambda_0}{\maru Z}_{ll_0}{}^{(l^1l_0^1,l^2l_0^2,l^3l_0^3)}
\nonumber\\
&=&
\left(-{\poon R}_{\frac{1}{2}-\frac{1}{2}}\right)^{\lambda+\lambda_0}\left({\poon R}_{\frac{1}{2},\frac{1}{2}}\right)^{\lambda-\lambda_0}
\left({\poon R}_-\right)^{l-l_0}{\maru Z}{}^{l^1l_0^1,l^2l_0^2,l^3l_0^3}\ . 
\eeq
Thus, multiplying the operator (\ref{40}) by (\ref{28}), we obtain the state generating operator with fifteen parameters.

Until the present, we have not contacted with the spherical tensor representation of our model 
with respect to the $su(2)$-algebras $({\wtilde R}_{\pm,0})$ and 
$({\wtilde S}_{\pm,0})$. 
Its explicit expression is omitted to show, but, we will discuss the basic idea for this problem. 
The linearly independent basis obtained in this paper is expressed in terms of twenty parameters, which, at least, five are 
related with the quantum numbers for the minimum weight state, i.e., $\rho$, $\rho^1$, $\sigma^1$, $\sigma^2$ and $\sigma^3$. 
Then, it may be important to investigate 
which parameters play a role of the quantum numbers or not. 
It is our final problem of this paper.

Application of the angular momentum coupling rule to the state (\ref{29}) leads us to the following: 
\bsub\label{40}
\beq
& &\mu\mu_0,\rho^1\rho_0^1\ \  \longrightarrow \ \ \mu\rho^1;\eta\eta_0\ , 
\label{40a}\\
& &\sigma^1\sigma_0^1,\sigma^2\sigma_0^2,\sigma^3\sigma_0^3 \ \ \longrightarrow \ \ \sigma^1\sigma^2 (\sigma^{12})\sigma^3;\sigma^{123}\sigma_0^{123}\ , 
\label{40b}
\eeq
\esub
\vspace{-0.9cm}
\bsub\label{41}
\beq
& &|\mu -\rho^1|\leq \eta \leq \mu+\rho^1 , 
\label{41a}\\
& &|\sigma^1-\sigma^2| \leq \sigma^{12} \leq \sigma^1+\sigma^2\ , 
\nonumber\\
& &|\sigma^{12}-\sigma^3| \leq \sigma^{123} \leq \sigma^{12}+\sigma^3\ . 
\qquad\qquad\qquad\ \ \ \ 
\label{41b}
\eeq
\esub
The inequalities (\ref{41a}) and (\ref{41b}) are related with the coupling rule for 
$({\wtilde R}_{\pm,0})$ and $({\wtilde S}_{\pm,0})$, respectively. 
If we note that $\rho^1$, $\sigma^1$, $\sigma^2$ and $\sigma^3$ denote 
the quantum numbers coming from the Casimir operator, $\mu$, $\eta$, $\sigma^{12}$ and $\sigma^{123}$ play a role 
of the parameters obeying the inequality (\ref{41}). 
Next, we consider the operator (\ref{39}). 
In this case, the coupling rule gives us 
\bsub\label{42}
\beq
& &\lambda\lambda_0,ll_0\ \  \longrightarrow \ \ \lambda l;\xi\xi_0\ , 
\label{42a}\\
& &l^1 l_0^1, l^2 l_0^2, l^3 l_0^3 \ \ \longrightarrow \ \ l^1 l^2 (l^{12}) l^3; l^{123} l_0^{123}\ , 
\label{42b}
\eeq
\esub
\vspace{-0.9cm}
\bsub\label{43}
\beq
& &|\lambda -l|\leq \xi \leq \lambda+ l , 
\label{43a}\\
& &|l^1-l^2| \leq l^{12} \leq l^1+l^2\ , 
\nonumber\\
& &|l^{12}-l^3| \leq l^{123} \leq l^{12}+l^3\ . 
\qquad\qquad\qquad
\label{43b}
\eeq
\esub
Under the relation (\ref{34}), the seven, $\lambda$, $\xi$, $l^1$, $l^2$, $l^{12}$, $l^3$ and $l^{123}$ have to be regarded as parameters obeying the 
inequality (\ref{43}). 
Finally, we consider ${\wtilde {\mib R}}^2$, ${\wtilde R}_0$, ${\wtilde {\mib S}}^2$ and ${\wtilde S}_0$ and, further, ${\wtilde R}$. 
Let the eigenvalues of ${\wtilde {\mib R}}^2$, ${\wtilde R}_0$, ${\wtilde {\mib S}}^2$ and ${\wtilde S}_0$ denote 
$r(r+1)$, $r_0$, $s(s+1)$ and $s_0$, respectively. 
For $({\wtilde {\mib R}}^2$, ${\wtilde R}_0)$ and $({\wtilde {\mib S}}^2$, ${\wtilde S}_0)$, we have the relations 
\bsub\label{44}
\beq
& &\xi\xi_0, \eta\eta_0\ \  \longrightarrow \ \ \xi\eta; r r_0\ , 
\label{44a}\\
& &l^{123} l_0^{123}, \sigma^{123} \sigma_0^{123} \ \ \longrightarrow \ \ l^{123} \sigma^{123}; s s_0\ , 
\label{44b}
\eeq
\esub
\vspace{-0.9cm}
\bsub\label{45}
\beq
& &|\xi -\eta| \leq r \leq \xi+ \eta\ , 
\label{45a}\\
& &|l^{123}-\sigma^{123}| \leq s \leq l^{123}+\sigma^{123}\ . 
\qquad\quad
\label{45b}
\eeq
\esub
At the present stage, we can summarize our discussion as follows: 
The eight symbols $\rho^1$, $\sigma^1$, $\sigma^2$, $\sigma^3$, $(r,r_0)$ and $(s, s_0)$ denote the quantum numbers 
and the eleven $\lambda$, $\mu$, $\xi$, $\eta$, $\sigma^{12}$, $\sigma^{123}$, $l^1$, $l^2$, $l^{12}$, $l^3$ and $l^{123}$ must 
be regarded as the parameters, totally the nineteen. 
However, in the above analysis, there are one point missing. 
We must take account of $\rho$ and $R$ (the eigenvalue of ${\wtilde R}$). 
After straightforward calculation, we can show that our linearly independent basis is a set of the 
eigenstates of ${\wtilde R}$ and $R$ can be expressed in the form 
\beq\label{46}
R=3(\mu-\lambda)+\frac{3}{2}(l^1+l^2+2l^3)-\rho\ .
\eeq
Then, we have to add $R$ and $\rho$ to the quantum numbers. 
Therefore, one of three parameters $\mu$, $\lambda$ and $(l^1+l^2+2l^3)/2$, for example, $\mu$, depends on the 
others if $R$ and $\rho$ are given.

Thus, we have a summary: 
In our linearly independent basis, $\rho$, $\rho^1$, $\sigma^1$, $\sigma^2$, $\sigma^3$, $r$, $r_0$, $s$, $s_0$ and $R$ denote 
the quantum numbers and $\sigma^{12}$, $\sigma^{123}$, $l^1$, $l^2$, $l^{12}$, $l^3$, $l^{123}$, $\xi$ and $\eta$ play a role 
of the parameters. 
Then, with the use of an appropriate method, for example, the Schmidt method, we can construct 
the orthogonal set for the $su(6)$-Lipkin model with arbitrary fermion numbers.

Including (I) and (II), this note has been devoted to the discussion on the Lipkin model, 
which belongs to the classical model in nuclear many-body theories. 
The authors tried to stress that there exist still some problems which remain 
unsolved until the present.

\section*{Acknowledgment}

Two of the authors (Y.T. and M.Y.) would like to express their thanks to 
Professor J. da Provid\^encia and Professor C. Provid\^encia, two of co-authors of this paper, 
for their warm hospitality during their visit to Coimbra in spring of 2015. 
The author, M.Y., would like to express his sincere thanks to Mrs K. Yoda-Ono for her cordial encouragement. 
The authors, Y.T., is partially supported by the Grants-in-Aid of the Scientific Research 
(No.26400277) from the Ministry of Education, Culture, Sports, Science and 
Technology in Japan.

\end{document}